# Magnonic Combinatorial Memory based on a network of coupled active ring circuits

Mykhaylo Balinskiy[1], Paulo Julio[1], Jeffrey Vargas[1], Diana Bisono Balaguer[1], Jacob Greenstein[2], and Alexander Khitun[1]

[1]*Department of Electrical and Computer Engineering, University of California - Riverside, Riverside, California, USA, 92521*

[2]*Department of Mathematics, University of California - Riverside, Riverside, California, USA, 92521*

Corresponding author: Alexander Khitun, email: akhitun@ece.ucr.edu

**Abstract:** Magnonic Combinatorial Memory (MCM) is a type of memory where the bits of information are encoded in the signal propagation paths in the network. In this work, we consider MCM based on the network of a coupled active ring circuit (ARC). Each circuit includes a broadband amplifier, a magnonic delay line, an adjustable frequency filter, an adjustable phase shifter, and a power detector. The coupling between the circuits is via spin waves propagating in the common delay line - ferrite film. There may or may not be auto-oscillations in the active ring circuits, depending on the combination of circuit parameters and circuit coupling. The address of MCM is defined as the combination of the states of the phase shifters and frequency filters, while the MCM state is defined as the presence/absence of the auto-oscillations. The coupling between the circuits is achieved by placing micromagnets on top of the ferrite film. The number of bits that can be encoded in the network increases quadratically with the number of coupled circuits. This scaling provides a fundamental advantage over conventional memory. We present experimental data obtained for three magnonic ARCs connected via a single-crystal yttrium iron garnet $Y_3Fe_2(FeO_4)_3$ (YIG) film. The data illustrate an example of encoding a 27-bit binary response pattern, corresponding to the 27 experimentally accessible phase combinations. The results demonstrate a robust operation of MCM with an On/Off ratio exceeding 30 dB at room temperature. The advantages and shortcomings of the proposed approach are discussed.

## Introduction

We are witnessing an unprecedented increase in the amount of data generated by humankind. The global volume of data created, captured, copied, and consumed in 2024 is estimated at 149 zettabytes [1]. The global data is expected to exceed 175 zettabytes (ZB) by 2025, according to the International Data Corporation [2]. It is projected to be an over 20 % increase in the global data volume every five years [3]. This exponential growth will shortly lead to the grand challenge for conventional storage systems, which may become unsustainable due to their limited data capacity, infrastructure cost, and power consumption [4]. For instance, there may not be enough silicon for flash-memory production by the year 2040 [5]. It stimulates the quest for novel memory devices with enhanced data storage density (i.e., the number of bits stored per area/volume/kilogram). *The number of bits stored in conventional memory devices is defined by the number of memory cells.* One cell can be in either two states (e.g., low resistance or high resistance) and store one bit of information. Over the past several decades, the enhancement in the storage density has been achieved by the miniaturization of the size of the memory cell. This trend is currently driving the research in nanometer-sized memory elements such as DNA-based [6] or sequence-defined macromolecules [7]. However, this approach may give only a temporary solution. It would be of great practical importance to develop a novel type of memory where the number of bits stored scales super linearly with the number of memory elements.

In our preceding works [8,9], we presented the concept of MCM and experimental data for the first working prototypes. The main idea of the combinatorial memory is to code information in the signal propagation paths in the mesh. The number of paths in the mesh is much larger compared to the number of elements in the mesh. It may be possible to increase the data storage density by assigning the bits of information to the paths between the elements rather than using individual elements. In the preceding works [8,9], we used a single multi-path magnonic active ring circuit (ARC). Here, we further evolve the idea of combinatorial memory to the system of coupled ARCs. It is expected to have more degrees of freedom to code information using the coupling between the circuits. The rest of the work is organized as follows. In Results, we describe the network model for combinatorial memory with coupled ARCs. We present the results of numerical modeling illustrating the phase map of the coupled circuits. The schematics of the prototype with three coupled active ring circuits, along with the obtained experimental data, are also given in Results. We discuss the advantages and limits of the proposed approach in Discussion.

## Results

### The principle of operation

Let us start with a single magnonic ARC as schematically shown in Fig.1(A). The circuit comprises electric and magnetic parts connected in series. The electric part includes a broadband amplifier G, an adjustable frequency filter, and an adjustable phase shifter $\Psi$, and a power detector. The magnetic part is a delay line - a waveguide made of material with low spin wave damping (e.g., YIG). Two microstrip antennas excite and receive spin waves through the waveguides. The detailed description of spin wave excitation and detection with micro-antennas can be found elsewhere [10]. The electric and magnetic parts are connected via coaxial cables. The signal circulating in the ring circuit exhibits a conversion from electromagnetic waves to spin waves and vice versa. The signal propagates as an electromagnetic wave in the electric part, and as a spin wave in the magnetic part. Spin waves propagating in the waveguide are much slower (e.g., $1\text{-}5 \times 10^4$ m/s) compared to the electromagnetic waves of the same frequency propagating in the coaxial cable. It provides a prominent phase shift $\Delta$ for even sub-millimeter-long waveguides. The magnitude of the shift depends on the spin wave dispersion (e.g., frequency-dependent). Signal conversion losses, as well as the spin wave damping, introduce signal damping $L$. The level of spin wave attenuation depends on many factors, including the quality of the ferrite film, the type of propagating spin wave, spin wave frequency, etc. There are two conditions for auto-oscillations to occur in ARC [11]:

$$G(V)+ L(f) \geq 0, \quad (1)$$

$$\Psi(V)+\Delta(f)=2\pi k, \text{ where } k=1,2,3, \ldots \quad (2)$$

where $G(V)$ is the amplifier gain in decibels, $L(f)$ is the signal attenuation in decibels, $\Psi(V)$ is the voltage-tunable phase shift, $\Delta(f)$ is the phase shift to the propagating signal provided by the element. The first equation (1) states the amplitude condition for auto-oscillations: the gain provided by the broadband amplifier should be sufficient to compensate for losses in the magnonic element. The second equation states the phase condition for auto-oscillations: the total phase shift for a signal circulating through the ARC should be a multiple of $2\pi$. In this case, signals come in phase every propagation round. *Only signal(s) propagating on the resonant frequency(s) that satisfy conditions (1) and (2) are amplified in the ARC.* It takes just a few rounds of signal circulation in the ring circuit untill the amplitude of the auto-oscillations reaches the maximum.

In general, there may be an infinite number of frequencies that satisfy the phase condition (1.2). For example, assuming the phase delay of the magnonic part to be a linear function of frequency $\Delta(f)$=$a$+$bf$, there is an infinite number of solutions for Eq.(2). This case is illustrated in Fig.1(B). The $\Delta(f)$ is shown as a function of $f$, where $\Delta(f)$ changes linearly from $0\pi$ to $2\pi$. Assuming $\Psi(V)$ to be set to $\pi$, there will be auto-oscillations for all the frequencies with $\Delta(f)$=$\pi$. The red dashed line in Fig.1(B) shows $\Delta(f)$=$\pi$. In our preceding work [12], 23 frequencies of auto-oscillations were simultaneously observed in one ARC. There is another observation one can make on the results in Fig.1(B). Figure 1(B) illustrates a simplified case in which the accumulated phase delay varies linearly with frequency. The horizontal dashed line represents one selected phase condition. Each intersection between the phase-delay curve and the dashed line corresponds to a discrete resonant frequency satisfying the Barkhausen phase condition. Consequently, an isolated ARC may support many discrete oscillation modes within the transmission band. The adjustable phase shifter allows these resonant modes to be shifted continuously, thereby selecting different oscillation frequencies."

Next, let us consider the two coupled ARCs as schematically shown in Fig.2(A). The coupling is through the element *J*, which provides an attenuation $L_{12}(f)$ and phase shift $\Delta_{12}(f)$ for the signals propagating from ARC-1 to ARC-2 and vice versa. For simplicity, we assume element *J* to be reciprocal, providing the same attenuation and phase shift for the signal propagating in both directions. Each isolated ARC supports auto-oscillations only at discrete resonant frequencies satisfying the amplitude and phase conditions (1) and (2). Since the phase accumulated during one round trip is a continuous function of frequency, these conditions can be satisfied by many (theoretically infinitely many) discrete resonant modes distributed over the transmission band. Figure 1(B) illustrates this concept using a simplified linear dependence of the accumulated phase on frequency, where the intersections with the selected phase condition correspond to the allowed oscillation frequencies. However, there exist phase combinations for which the coupling between the active ring circuits produces destructive interference between multiple signal propagation paths. As a consequence, the effective loop gain falls below the oscillation threshold, resulting in suppression of self-sustained oscillations (oscillation or amplitude death).This formalism can be rewritten in terms of per round-trip $n$ instead of continuous time $t$ as follows [13]:

$$A_1^{(n+1)}=G_1\left(\left|A_1^{(n)}\right|^2\right)e^{-L_1}e^{i\Phi_1}A_1^{(n)}+JA_2^{(n)}$$
$$A_2^{(n+1)}=G_2\left(\left|A_2^{(n)}\right|^2\right)e^{-L_2}e^{i\Phi_2}A_2^{(n)}+JA_1^{(n)} \quad (3)$$

where $G_j\left(\left|A_j^{(n)}\right|^2\right)$ is the saturating loop gain**,** $L_j$ is total attenuation per round-trip, $\phi_j$ is the total phase shift per round-trip $\phi_j$=$\Psi_j(V)$+$\Delta_j(f)$. Linearizing these maps or taking the limit of small round-trip time directly leads back to the continuous-time equations above. Assuming the two rings have the same gain and the same coupling, Eq.(3) can be further simplified to the following discrete round-trip equation in the matrix form:

$$\begin{pmatrix} A_1^{(n+1)} \\ A_2^{(n+1)} \end{pmatrix} = \underbrace{\begin{pmatrix} Ge^{i\Phi_1} & J \\ J & Ge^{i\Phi_2} \end{pmatrix}}_{T} \begin{pmatrix} A_1^{(n)} \\ A_2^{(n)} \end{pmatrix} \quad (4)$$

Let us take the eigenvalues of $T$ to be $\lambda$. The oscillations happen when the largest $|\lambda|$ reaches 1. At the threshold, at least one mode satisfies $|\lambda|$=1.

In Figure 2(B), we present the results of numerical modeling by Eq.(4), showing the example phase map for the two coupled oscillators. The simulations are accomplished for $G$=1, and $J$=0.3. The two phases were changed from 0 to $2\pi$: $\Psi_1 \in [0, 2\pi]$, $\Phi_2 \in [0, 2\pi]$. The calculation procedure is the following. For each pair of $(\Psi_1, \Psi_2)$ the matrix $T$ is built. The two eigenvalues $\lambda_{1,2}$ are computed. We marked that point as oscillatory if $max(|\lambda_1|, |\lambda_2|)>1$. In Fig. 2(B), the X and Y axes are $\Psi_1$ and $\Psi_2$, respectively. The red region (value 1): phase combinations that lead to oscillations (growth). The blue region with two “islands” (value 0) corresponds to the case when the largest eigenvalue magnitude drops below 1 (no oscillation). The reason for the no oscillation is the following. The coupling between the circuits opens additional signal propagation paths. The interference between the waves propagating on the different paths may be destructive, leading to the violation of the amplitude or phase conditions. For example, let us consider the paths for signal propagation starting from the point marked ① to the point marked ②. One path is through $\Psi_1$ and $L_1\angle\Delta_1$. The second path is through $\Psi_1$, $L_{12}\angle\Delta_{12}$, and $L_2\angle\Delta_2$. The third path is through $\Psi_1$, $L_{12}\angle\Delta_{12}$, $\Psi_2$, and $G_2$. The wave coming to the point ② is a superposition of three waves:

$$\frac{A_{at\,2}}{A_{at\,1}}=\Xi=[L_1\angle(\Delta_1+\Psi_1)+L_{12}\ L_2\angle(\Psi_1+\Delta_{12}+\Delta_2)+L_{12}\ G_2\angle(\Psi_1+\Delta_{12}+\Psi_2)]. \quad (5)$$

The amplitude and the phase conditions for auto-oscillation are the same as for the single ARC:

$$\mathrm{Re}[\Xi]>1,$$
$$\mathrm{Im}[\Xi]=2\pi k,\ \text{where}\ k=1,2,3,\dots \quad (6)$$

The waves propagating along the different paths interfere destructively at the amplifier input, reducing the effective loop gain below unity. Consequently, the system enters a regime of oscillation (amplitude) death in which self-sustained oscillations cannot be established." *The appearance of a no-oscillation region in the phase map is the most important difference between a single ARC and the two coupled ARCs. Some of the phase combinations $\Psi_1$ and $\Psi_2$ will lead to the self-oscillations, and others will not.*

This approach can be extended to the system of three coupled ARCs. The formalism for the three-coupled circuit is similar to the one shown above for the two coupled oscillators. The system of equations for the three-coupled oscillators can be given as follows:

$$A_1^{(n+1)}=Ge^{i\phi_1}A_1^{(n)}+J_{12}A_2^{(n)}+J_{13}A_3^{(n)}$$
$$A_2^{(n+1)}=J_{21}A_1^{(n)}+Ge^{i\phi_2}A_2^{(n)}+J_{23}A_3^{(n)}$$
$$A_3^{(n+1)}=J_{31}A_1^{(n)}+J_{32}A_2^{(n)}+Ge^{i\phi_3}A_3^{(n)} \quad (7)$$

The coupling between three ARCs leads to a more complicated phase map with oscillation and no oscillation. The phase maps can be engineered by changing the coupling coefficients as well as changing the level of amplification in ARCs. A collection of phase maps obtained for the different combinations of circuit parameters (e.g., gain, loop phase) and circuit-to-circuit coupling (same coefficients or different, reciprocal and not-reciprocal) is presented in the Supplementary Materials.

In general, the system of $N$-coupled ARCs can be described as follows:

$$A^{(n+1)}=TA^{(n)}, \quad (8)$$

where $T_{ll}=K_l=G_l e^{i\phi_l}, 1 \leqslant l \leqslant N$ are the per-circuit complex round-trip factors, and $T_{kl}=J_{kl}, 1 \leqslant k \neq l \leqslant N$, are the coupling matrix elements. The system is described by the $N$-dimensional phase map where some of the phase combinations lead to the auto-oscillations, and others do not. It is important to note that the per-circuit complex round-trip factors $K_l$, and the coupling matrix elements $J_{kl}$ are frequency dependent. There may be more than one frequency of auto-oscillations corresponding to one phase combination. Also, the level of amplification in each ARC affects not only the auto-oscillation conditions in the circuit but also the coupling between the other circuits. The correlation between the phase combination and circulation power (i.e., auto-oscillation or no auto-oscillation) is the essence of the combinatorial memory operation. We define memory address and state as follows:

$Memory\ address$: $\{\Psi_1,..\Psi_N, f_1,..f_N, G_1,..G_N\}$, $\quad Memory\ state$: $\{P_1,..P_N\}$

where $\Psi_j$ is the voltage-adjustable phase shifter, $f_j$ is the passband frequency (i.e., defined by the frequency filter) per circuit, $G_j$ is the amplification level oscillation, $P_j$ is the state of the power detector (e.g., state 1 if $P_j$>$P_{ref}$ and state 0, otherwise, where $P_{ref}$ is some reference power level.

The schematics of MCM with four coupled ARCs are shown in Fig.3. Each circuit includes a non-linear broadband amplifier, an adjustable frequency filter, an adjustable phase shifter, and a power detector. The coupling between the circuits is via spin waves propagating in the common delay line - ferrite film. There are eight microstrip lines shown on the sides of the film. The microstrips on the left side, marked as 1,2,3, and 4, are to excite spin waves. The microstrips marked as 5,6,7, and 8 are to convert spin waves into an AC voltage. Because the active ring circuits are coupled through the common ferrite film, signals may propagate along multiple closed-loop trajectories involving different ARCs before returning to their original circuit. The superposition of these multiple propagation paths determines the amplitude and phase conditions for self-sustained oscillation. The number of possible paths scales exponentially with the number of coupled ARCs. In turn, it leads to a large number of possible correlations between the memory state and memory addresses. One phase combination may lead to more than one auto-oscillation frequency in the network. That is the reason for using the set of frequency filters that allows us to extract information for one specific frequency per ARC at a time. The output of the device is the *N*-bit number defined by the states of the power sensors. It should be noted that the power in the output #6 may be in the state On even for the low or no-amplification provided by the amplifier $G_1$ due to the collective work of other amplifiers.

The key question is the following. *How many bits can one store in the system of N-coupled ARCs?* The dynamics of the system is described by the set of *N* equations (see Eq.(8), where each equation contains *N* terms: $(N\text{-}1)$ coupling matrix elements $J_{ij}$ and per-circuit complex round-trip factors $K_j$. In turn, the terms consist of the variable and non-variable parts. For instance, we change the voltage-adjustable phases $\{\Psi_1, .. \Psi_N\}$, combination of the frequency filters $\{f_1, .. f_N\}$, and the amplification levels of the amplifiers $\{G_1, .. G_N\}$. These are the variables reserved for the memory addresses aimed at reading out information. The information is encoded in the coupling coefficients $J_{ij}$. The maximum number of bits that can be encoded $B$ can be calculated as follows:

$$B = N \cdot (N\text{-}1) \cdot n_{JL} \cdot n_{J\Psi} \cdot n_{f,} \quad (9)$$

where $N \cdot (N\text{-}1)$ is the number of coupling coefficients, $n_{JL}$ is the number of states the amplitude (i.e., the attenuation) for the coupling coefficient, $n_{J\Psi}$ is the number of phase states per element, and $n_f$ is the number of the operational frequencies. The number of bits scale proportionally to $N \cdot (N\text{-}1)$ according to Eq.(9). This is the key statement that can be rigorously mathematically proved. Let $A$ be an $N \times N$ matrix with complex entries. By definition, the spectral radius $\rho(A)$ of $A$ is $\max\{|\lambda|: \lambda \in \mathbb{C} \text{ is an eigenvalue of } A\}$. It is well-known that $\rho(A) \geq \frac{1}{N}|Tr(A)|$. Indeed, let $\lambda_1, \dots \lambda_N \in \mathbb{C}$ be eigenvalues of $A$ (counting multiplicities). Then $|Tr(A)| = |\sum_{1 \leq k \leq N} \lambda_k| \leq \sum_{1 \leq k \leq N} |\lambda_k| \leq N\rho(A)$.

In order to obtain auto-oscillations in our system, we need to ensure that $\rho(T)$>1. In view of the above formula, a sufficient condition for that can be obtained as follows. Since the diagonal entries of $T$ are of the form $K_l$=$G_l e^{i\phi_l}$, $1 \le l \le N$, we have

$$|Tr(T)|^2 = \sum_{1\le k,l\le N} G_k\, G_l e^{i(\phi_k-\phi_l)} = \sum_{1\le k\le N} {G_k}^2 + 2\sum_{1\le k<l\le N} G_k\, G_l cos(\phi_k-\phi_l) \qquad (10)$$

Let 0<$R \leqslant N$ be the number of $1 \leqslant k \leqslant N$ such that $G_k$>0. If (a) $G_k \ge \sqrt{2}\frac{N}{R}$ if $G_k$>0; (b) $|\phi_k-\phi_l| \le \frac{\pi}{3}$ for all $1 \le k$<$l \le N$ such that $G_k G_l \ne 0$

$$|Tr(T)|^2 \ge \frac{2N^2}{R} + \frac{(R\text{-}1)N^2}{R} = \frac{(R+1)N^2}{R} > N^2, \qquad (11)$$

which forces $\rho(T)$>1.

If we assume that all entries in the matrix $T$ are real, then by the same argument $\rho(T)$>1 provided that $G_k$>$\frac{N}{R}$ for all $1 \leqslant k \leqslant N$ such that $G_k$>0. Thus, we “lose” the free usage of diagonal entries of our matrix, since they are meant to control its spectral radius. Yet, we can still store some using diagonal entries since we can choose which of them are zero and which are not. For example, we only consider the situation when $R \geqslant N/2$, then the number of choices of possible locations of non-zero entries grows as $2^{N/2}$ , and so we get additional $N/2$ bits. Let us return to the rest of the matrix. Assume for simplicity that all entries are real (which, in particular, ensures that the phase shift is a multiple of $2\pi$). We encode information by selecting off-diagonal entries in the matrix in such a way that, when we put a signal at just one entry #$k$ with power level $P$, the output levels at exit $l \ne k$ is a prescribed fraction $\left(r_{lk}/n_{JL}\right)^2 P$, where $0 \leqslant r_{lk}$<$n_{JL}$. This is achieved by choosing the coupling constant $J_{lk}$=$r_{lk}/n_{JL}$. The information on the $k$th exit is whether the signal level is non-zero and is determined by the value of $G_k$. Thus, the total number of bits that can be stored in this model is $log_2\left(n_{JL}\right)N(N-1)$+ $N/2$ =$O(N^2)$.

If the matrix $T$ is symmetric the situation changes a bit, but the growth rate of the number of bits is still quadratic. First of all, in that case, we can use a weaker sufficient condition to guarantee that $\rho(T)$>1. Indeed, as for $T$ symmetric $\rho(T) \geqslant \|Tv\|$ for any vector $v$ with $\|v\|$=1 where $\|\cdot\|$ is the usual Euclidean norm, we immediately obtain that $\rho(T) \geqslant \sqrt{\sum T_{kl}^2} \geqslant \max_{1\leqslant k\leqslant N}|T_{kl}|$ for any $1 \leqslant l \leqslant N$ which in turn implies that $\rho(T) \geqslant \max_{1\leqslant k,l\leqslant N}|T_{kl}|$. Thus, as $|T_{kl}|$<1, $k \ne l$, it suffices to have one diagonal entry $T_{kk}$=$G_k$>1, $1 \leqslant k \leqslant N$. This can be improved even further if we assume that the $l$th column of $T$ (and, since $T$ is symmetric, the $l$th row) for some $1 \leqslant l \leqslant N$ does not contain any zero entries. In that case, $\rho(T) \geqslant \sqrt{\sum T_{kl}^2} \geqslant \sqrt{N}\min_{1\leqslant k\leqslant N}|T_{kl}|$. Therefore, $\rho(T)$>1 provided that *every entry in some column (and hence a row)* is strictly greater than $1/\sqrt{N}$ by absolute value.
The symmetry means that we can choose only $N\ (N+1)/2$ matrix entries arbitrarily. The factor ½ does not affect the growth rate and can be compensated by the number of possible states at exits anyway.

The area of the MCM $S$ scales linearly with the number of coupled circuits $S$=$N \cdot S_0$, where $S_0$ is the area of a single ARC with the area of the common ferrite film included. Thus, the data storage density $B/S$ scales as

$$B/S \sim N/S_0. \quad (12)$$

According to Eq.(12), the data *storage density increases* with the number of coupled circuits *N*, that provides a fundamental advantage over the conventional memory where the data storage density remains constant regardless of the number of data storage elements. In the derivation of Eq.(12), it was assumed that we can *independently control all the coupling coefficients* (i.e., the attenuation, the phase shift, at all selected frequencies) for all coupling coefficients. Though it is theoretically possible, there are technological constraints to be further considered in the Discussion. Next, we present experimental data showing the operation of the prototype with three coupled ARCs.

**Prototype structure and experimental data**

The cross-sectional view of the MCM device is shown in Fig. 4(A). It consists from the bottom to the top of a permanent magnet, Printed Circuit Board (PCB) with six antennas, Gadolinium Gallium Garnett (GGG) substrate, and a YIG film. The permanent magnet is a pair of commercially available NdFeB magnets (model BY0Y04 by K&J Magnets, Inc.) with the dimensions of 2.0" × 2.0" × 0.25". It is aimed to create a constant bias magnetic field. The bias field is about 375 Oe and directed in-plane on the YIG film surface. The ferrite film is made of YIG grown by liquid epitaxy on a GGG substrate. The film is not patterned. The thickness of the film is 42 µm. The saturation magnetization is close to 1750 G, the dissipation parameter (i.e., the width of the ferromagnetic resonance) ΔH = 0.6 Oe measured at 3 GHz. The YIG layer is placed on top of the PCB with antennas. The thickness of the GGG substrate is reduced by polishing to 0.3 mm from the initial 0.5 mm. A micro-magnet is to be placed on top of the YIG layer. There are six micro antennas fabricated on the PCB. The antennas are numbered 1 to 6 as shown in Fig.4(B). The characteristic size of the antenna is 6 mm in length and 0.15 mm in width. These antennas are used as the input/output ports for spin wave excitation/detection.

The schematics of the experimental setup are shown in Fig.5. There are three active ring circuits sharing the same ferrite film. We differentiate the circuits by the pair of spin wave generating and spin wave-receiving antennas. ARC-1 uses antenna #1 to excite spin waves and antenna #4 to convert spin waves into AC voltage. ARC-1 consists of the phase shifter $\Psi_1$, an amplifier $G_1$, a frequency filter $f_1$, and a directional coupler $P_{14}$. ARC-2 uses antenna #6 to excite spin waves and antenna #3 to convert spin waves into AC voltage. ARC-2 consists of the phase shifter $\Psi_2$, an amplifier $G_2$, a frequency filter $f_2$, and a directional coupler $P_{52}$. ARC-3 uses antenna #5 to excite spin waves and antenna #2 to convert spin waves into AC voltage. ARC-3 consists of the phase shifter $\Psi_3$, an amplifier $G_3$, a frequency filter $f_3$, and a directional coupler $P_{63}$. The directional couplers are used to check the level of the circulating power. The frequency filters are commercially available YIG-sphere-based voltage-tunable bandpass filters produced by Micro Lambda Wireless, Inc, model MLFD-40540. Experimental data on the filter characteristics can be found in the Supplementary Materials. The phase shifters are commercially available shifters from ARRA, model 9418 A. The accuracy of the phase control is about $1^0$ in the operational frequency range. The directional couplers are KRYTAR, model 1820. The broadband amplifiers are produced by Mini-Circuits, model ZX60-83LN-S+. The filters are set to the following central frequencies $f_1$= 1.614 GHz, $f_2$= 1.838 GHz, and $f_3$= 1.720 GHz, respectively. This is one of the many possible frequency combinations that have been chosen for experiments due to the minimum losses.

The first set of experiments was accomplished without a magnet on top of the ferrite film. In Fig.6, there is shown the correlation between the phase triplets and oscillation powers in the coupled circuits. In Fig.6(A), the adjustable phase shifter in ARC-1 was set to $\Psi_1$=$0\pi$. The phase shifters $\Psi_2$ and $\Psi_3$ in ARC-2 and ARC-3 were independently set to $0\pi$, $2\pi/3$, and $4\pi/3$, respectively. In Fig.6(B), the adjustable phase shifter in ARC-1 was set to $\Psi_1$=$2\pi/3$. The phase shifters $\Psi_2$ and $\Psi_3$ in ARC-2 and ARC-3 were independently set to $0\pi$, $2\pi/3$, and $4\pi/3$, respectively. In Fig.6(C), the adjustable phase shifter in ARC-1 was set to $\Psi_1$=$4\pi/3$. The phase shifters $\Psi_2$ and $\Psi_3$ in ARC-2 and ARC-3 were independently set to $0\pi$, $2\pi/3$, and $4\pi/3$, respectively. There are 9 markers in each plot showing the oscillation power detected through the directional coupler $P_{14}$ in dBm. The power of – 90 dBm corresponds to the noise level (no auto-oscillations). The On state corresponds to the power above -60 dBm. In Fig.6(D), we combined all three plots. It demonstrates the 3D phase map where some of the phase combinations lead to self-oscillations and others do not. Next, we repeated the experiments with a micro-magnet placed close to antenna #6 as shown in Fig. 7(A). The volume of the NdFeB micro-magnet is about 0.05 mm$^3$. The experimental data obtained with (blue markers) and without (red markers) magnet is presented in Fig.7(B). We plotted only phase triplets that lead to the auto-oscillations where the power taken through the $P_{14}$ coupler is above -60 dBm.

## Discussion

There are several observations we can make based on the obtained experimental data. (i) The experimental data show the practical feasibility of engineering phase maps for the system of coupled ARCs where some of the phase combinations lead to the auto-oscillations, and others do not. We want to stress that the data presented in Figs.6 and 7 are collected with the fixed level of amplification and a fixed position of the frequency filters for all three ARCs. Only the adjustable phase shifters were changed over the three selected phase states. The change in the circulating power is only due to the change in the phase condition (2). (ii) Placing a magnet on top of the ferrite film may significantly affect the coupling between the coupled magnonic ARCs. The data presented in Fig.7(B) demonstrate an example of using just one magnet to encode 27 bits of information. (iii) There is a prominent dependence on the oscillation power for different phase combinations. The On/Off ratio (i.e., oscillations/no oscillations) exceeds 30 dB at room temperature (see Figs.6 and 7). The high On/Off ratio is a must for building robust memory devices. The achieved high On/Off is due to the utilization of the phase condition 1.2 and the interplay between the ARCs. The circulating power is almost the same (+- 3dB) for all the phases in the not-coupled ARCs. Read reliability is one of the most important performance metrics for any memory technology. In the present work, the measured ON/OFF ratio exceeding 30 dB provides a large sensing margin between the oscillatory ("1") and non-oscillatory ("0") states, thereby reducing the probability of read errors caused by thermal fluctuations, amplifier noise, and detector uncertainty. The read error rate depends not only on the ON/OFF ratio but also on the stability of the oscillation power, phase noise, detector sensitivity, and environmental variations. A comprehensive statistical characterization of the bit-error rate, including repeated read cycles under different operating conditions and temperatures, is beyond the scope of the present proof-of-concept demonstration and will be the subject of future work.

It should be noted that the collected experimental data show only a small fraction of the bits that can be stored in the three coupled ARCs. Only 27 addresses corresponding to $3^3$=27 phase combinations were used while the levels of amplification for all circuits were fixed. The data is collected only for one frequency combination. Conservatively taking two levels of amplification (e.g., max amplification and no amplification) $n_{JL}$=2, three phases $n_{J\Psi}$=3, and two operational frequencies $n_f$=2, one obtains $B$= 72 using Eq.(9). The storage density of the prototype can be estimated as 72/(20 mm$^2$) ≈2.3 kbit/in$^2$. Modern magnetic hard drives commercially available today have storage densities exceeding 1 terabit per square inch (1 Tbit/in$^2$) [14]. Nevertheless, there is a big room for MCM data storage enhancement by reducing the size of circuit elements (e.g., spin wave generating/receiving antennas), increasing the number of states that can be recognized for the coupling coefficients, increasing the number of operational frequencies, and, most importantly, the number of coupled ARCs. For instance, scaling down the size of the device to 20 μm$^2$, using 10 operational frequencies, and increasing the number of coupled ARCs to 30 would increase the storage density of MCM above 1 Tbit/in$^2$. Further scaling down to sub-micrometer feature size together with the increase in the number of coupled ARCs will provide a drastic enhancement in the data storage density compared to any existed or proposed memory.

The independent control of all coupling coefficients is the main technological challenge. The maximum data storage density is estimated assuming *the individual control of all coupling coefficients*. For instance, it presumes a non-reciprocal coupling $J_{ij} \neq J_{ji}$. There are multiple works on non-reciprocal spin wave transport [15-18]. However, the practical implementation of the non-reciprocity is quite challenging. The number of bits stored will be reduced by half for using reciprocal coupling. The difference in the strength of coupling between the ARCs is another grand challenge. As one can see from Fig.3, the coupling between the nearest ARCs (e.g., 1 and 2, 2 and 3) will be stronger (i.e., less attenuation) compared to the coupling between the distant circuit ARCs (e.g., 1 and 4). This problem can be solved by using duplicate input/output ports for a single ARC. It can also be resolved by increasing the level of amplification in all ARCs to compensate for losses, even for the most distant ARCs. In either case, the solution would require either additional resources or an increase in the power consumption.

Read-only memory (ROM) is the most promising application of the proposed combinatorial memory. In MCM, the stored information is encoded by engineering the coupling coefficients between the coupled active ring circuits. In the present prototype, this is achieved by positioning permanent micromagnets above the ferrite film. Determining the appropriate magnet configuration for a desired data string is a coding problem whose mathematical framework has been described in Ref. and is not repeated here. During read-out, the desired memory address is selected by setting the phase shifters $\Psi_j$ , amplifier gains $G_j$, and frequency filters $f_j$ of the individual ARCs. Since ROM devices are programmed infrequently, programming time is generally less critical than read latency. The stored information is non-volatile because it is defined by the permanent magnetic configuration and therefore does not require electrical power for retention. Power is required only during read-out, when the coupled ARCs evolve to the self-sustained oscillatory state used for data retrieval.

Modern Spin-Transfer Torque Magnetic Random Access Memory (STT-MRAM) and related magnetic memories exhibit read latencies on the order of ~5–20 ns. Reported read energies are typically in the range of ~10–100 fJ per bit, depending on junction resistance, sensing circuitry, and technology node [19-21]. The read-out time of MCM is defined by the number of rounds required for the system to reach amplitude saturation. This time was about 1 ms in our preceding works based on the mm-scale structures [9,22]. This time can be reduced by about 10-100 times by scaling down the size of the spin wave waveguides. The energy consumption of MCM is defined by the level of amplification and time required to establish self-sustained auto-oscillations. The power consumed by the amplifiers was about 1 mW[9,22]. The estimated energy consumption for one bit is currently 1μJ and, potentially, may be reduced by one to two orders of magnitude. Although self-sustained oscillations exist only during the read-out process, the stored information itself is not encoded in the oscillation amplitudes but in the coupling coefficients between the active ring circuits. In the present prototype, these coupling coefficients are defined by the positions of permanent micromagnets placed above the ferrite film. Since the magnetic configuration remains unchanged after the power is turned off, the stored information is retained without electrical power. When power is restored, the same coupling configuration reproduces the same phase map and, consequently, the same memory content. Thus, the proposed MCM architecture is inherently non-volatile. Based on these estimates, one can conclude that MCM cannot compete with traditional memory in speed and read-out power consumption. Data storage density is the most important advantage of MCM over conventional memory.

There is another fundamental advantage of MCM to be mentioned. It makes it possible to perform a parallel database search. A number of phase combinations can be checked in parallel by removing (i.e., not using) the frequency filters. The device will evolve to a state with at least one ARC in the On state if there is just one frequency or combination of frequencies that satisfy the amplitude and phase conditions. Wave superposition is the origin of the parallel MCM search, where waves of different frequencies propagate and are amplified or not-amplified at the same time. The power of parallel database search using classical wave superposition has been considered in [23]. This appealing property of MCM with coupled ARCs deserves special consideration.

There are some critical comments on the material presented in this work. The collected experimental data showing the change of the phase map due to the presence of a magnet is not supported by the results of numerical simulations or theoretical analysis. The presented experimental data in Fig.7 show only an example of controlling the coupling between the ARCs using just one magnet. A more detailed study on the relation between the magnet position and coupling coefficients can be found in Ref. [9]. The relation between the position of a magnet on top of a ferrite film to the S-parameters of a multi-port network was demonstrated. A deeper study is needed to prove that *any* of the desired phase maps can be engineered by placing magnet(s) on top of the common ferrite film. An important challenge for practical implementation of MCM is the inverse programming problem: determining the magnetic configuration required to generate a prescribed binary response pattern. While the present work demonstrates that controlled modifications of the magnetic landscape alter the coupling matrix and the corresponding memory response, an efficient deterministic programming algorithm has not yet been developed. Solving this inverse-design problem for large networks, without relying on exhaustive search, will likely require numerical optimization or machine-learning-assisted approaches and constitutes an important direction for future research.

In the present work, the feasibility of enhancing the data storage density by exploiting a number of frequency bands. The frequency band of the filters used in this work is 15 MHz, while the width of the spin wave transmission band in the film is about 2 GHz. Meaning that more than 100 frequency channels can be utilized. However, the ability to code information is directly related to the ability to control coupling coefficients (i.e., to have different phase shift/attenuation) at different frequencies. The number of bits stored would not increase by using the identical coupling (e.g., having a linear dispersion as shown in Fig.1(B)). Fortunately, spin waves have a great variety of dispersion characteristics, where the phase velocity depends on the magnitude, the direction of the applied magnetic field, and frequency [24]. A comprehensive investigation of fabrication tolerances, statistical distributions of coupling coefficients, calibration procedures, and process-induced variability will be an important subject of future work.

Another important practical consideration is the correlation between coupling coefficients. Since all active ring circuits are coupled through the same ferrite film, modifying the magnetic configuration in one region generally changes several coupling coefficients simultaneously.

Therefore, the encoded bits cannot, in general, be regarded as completely independent. The storage-density estimates presented in this work represent an upper theoretical limit corresponding to independently controllable coupling coefficients. In practical implementations, the number of independently programmable bits will depend on the number of independent degrees of freedom available for engineering the magnetic landscape. Determining these limits and developing optimization algorithms for programming the coupling matrix constitute important directions for future research. In part, this problem has been addressed in Ref.[25].

Coupled-oscillator networks have been explored both as data-storage (associative memory) media—where stored patterns correspond to stable phase-locked/synchronized states—and as data-processing hardware for tasks such as pattern recognition and combinatorial optimization. For storage, phase-based oscillatory neural networks can store and retrieve patterns via synchronization with specific phase relations (e.g., PLL neural networks)[26], and the capacity limits of oscillatory associative-memory networks have been analyzed[27]. Hardware-oriented associative-memory implementations based on arrays of coupled oscillators have also been proposed and evaluated for pattern recognition [28]. For data processing, coupled oscillators are widely used as analog compute engines; for example, an experimentally demonstrated weighted Ising machine [29] built from coupled nonlinear LC oscillators solves optimization instances (MAX-CUT) and highlights rapid convergence in a few oscillator cycles, while oscillator-array models provide circuit-level frameworks for pattern recognition/classification. Spin Waves in Magnetic Film Feedback Rings have been studied in a number of works [30]. This work describes a novel approach to data storage using magnonic coupled oscillators. It aims to benefit from the advantages of magnetic memory: non-volatility and long endurance time, and shows a way to fundamentally increase data storage density.

In summary, we considered a memory device based on coupled magnonic ARCs. The coupling is via spin waves propagating in the common ferrite film. The operation of the memory is based on the correlation between the circuit parameters and the presence/absence of the self-sustained auto-oscillations. The information is encoded in the circuit-to-circuit coupling that can be controlled by placing magnets on top of the ferrite film. We presented experimental data showing the operation of MCM based on three ARCs coupled through the single-crystal yttrium iron garnet $Y_3Fe_2(FeO_4)_3$ (YIG) film. It is shown an example of encoding a 27-bit binary response pattern, corresponding to the 27 experimentally accessible phase combinations, using a single micromagnet. The On/Off ratio exceeds 30 dB at room temperature. The number of bits that can be stored in MCM scales quadratically with the number of coupled ARCs. It provides a fundamental advantage over the existing memory devices. Overall, the obtained results are in favor of the magnonic combinatorial memory approach that may provide a fundamental increase in the data storage density compared to conventional memory devices.

## Methods

### Device Fabrication

The core of the device is made of a single-crystal $Y_3Fe_2(FeO_4)_3$ film. The film was grown on top of a (111) Gadolinium Gallium Garnett ($Gd_3Ga_5O_{12}$) substrate using the liquid-phase epitaxy technique.  The thickness of the film is 42 μm. The saturation magnetization is close to 1750 G, the dissipation parameter (i.e., the width of the ferromagnetic resonance) $\Delta H$ = 0.6 Oe. The bias magnetic field is provided by the permanent magnet made of NdFeB.

### Measurements

The excitation and detection of spin waves in the ferrite film were accomplished by six short-circuited antennas. The antennas are connected to a programmable network analyzer (PNA) Keysight N5241A. The filtering is by the commercially available filters produced by Micro Lambda Wireless, Inc, model MLFD-40540.

### Data availability

All data generated or analyzed during this study are included in this published article and the Supplementary Materials.

**Author contributions**

M.B. built the prototype and accomplished experiments. P.J. assisted with data acquisition. J.V. and D.B. assisted with data arrangement. J.G. provided the mathematical analysis. A.K. conceived the idea of combinatorial memory. All authors wrote and reviewed the manuscript.

**Competing financial interests**

The authors declare no competing financial or non-financial interests

**Acknowledgments**

This work was supported by the National Science Foundation under grant # 2423929 and is supported in part by funds from federal agency and industry partners as specified in the Future of Semiconductors (FuSe) program.

**Figure Legends**

**Figure 1:** Single ARC. (A) Schematics of the magnonic active ring circuit that consists of a broadband amplifier $G(V)$, a spin wave waveguide - delay line, and a voltage-tunable phase shifter $\Psi(V)$. The two antennas on top of the waveguide are aimed to generate spin waves and convert spin waves into AC voltage. Spin waves propagating in the waveguide accumulate amplitude change $L(f)$ and phase shift $\Delta(f)$. (B) Results of numerical simulations showing the phase shift accumulated during spin wave propagation in the waveguide in the case of a linear dispersion. The red dashed line shows one selected phase shift that satisfies the phase condition for auto-oscillations. There may be infinite frequencies that satisfy the phase condition for auto-oscillations.

**Figure 2:** Two coupled ARCs. **(**A) Schematics of the circuit with two coupled ARCs. The coupling is via the element *J* which provides $L_{12}(f)$ attenuation and $\Delta_{12}(f)$ phase shift to the propagating signals. (B) Results of numerical modeling showing the phase map of the circuit. The red color corresponds to the phase combinations satisfying the phase condition of auto-oscillation. The blue color shows the regions in the phase map where self-oscillations are not possible due to the destructive signal interference.

**Figure 3:** Schematics of MCM with four coupled ARCs. Each ARC includes a non-linear broadband amplifier, an adjustable frequency filter, an adjustable phase shifter, and a power detector. The coupling between the circuits is via spin waves propagating in the common delay line - ferrite film. There are eight microstrip lines shown at the sides of the film. The microstrips on the left side, marked as 1,2,3, and 4 are to excite spin waves. The microstrips marked as 5,6,7, and 8 are to convert spin waves in AC voltage. The output of the device is the *N*-bit number defined by the states of the power sensors.

**Figure 4:** Device under study. (A) The cross-sectional view of the device. It consists from the bottom to the top of a permanent magnet, Printed Circuit Board (PCB) with six antennas, and a YIG film fabricated on the Gadolinium Gallium Garnett (GGG) substrate. The thickness of the YIG film is 42 μm. The saturation magnetization is close to 1750 G, the dissipation parameter (i.e., the half-width of the ferromagnetic resonance) ΔH = 0.6 Oe measured at 3 GHz. (B) The photo of the PCB with six antennas. The characteristic size of the antenna is 6 mm in length and 0.15 mm in width. These antennas are used as the input/output ports for spin wave excitation/detection. The antennas numerated 1-6 to define the ARCs.

**Figure 5:** Schematics of the experimental setup. There are three active ring circuits sharing the same ferrite film. Each circuit consists of a voltage-adjustable frequency filter, voltage-adjustable phase shifter, directional coupler, and a non-linear broadband amplifier. There are two antennas per circuit for spin wave generation/detection. ARC-1 uses antenna #1 to excite spin waves and antenna #4 to convert spin waves into AC voltage. ARC-2 uses antenna #6 to excite spin waves

and antenna #3 to convert spin waves into AC voltage. ARC-3 uses antenna #5 to excite spin waves and antenna #2 to convert spin waves into AC voltage.

**Figure 6:** Collection of experimental data showing the correlation between the phase triplets and the power circulating in ARC-1. (A) $\Psi_1$=$0\pi$, $\Psi_2$ and $\Psi_3$ are independently set to $0\pi$, $2\pi/3$, and $4\pi/3$. respectively. (B) $\Psi_1$=$2\pi/3$, $\Psi_2$ and $\Psi_3$ are independently set to $0\pi$, $2\pi/3$, and $4\pi/3$, respectively. (C) $\Psi_1$=$4\pi/3$, $\Psi_2$ and $\Psi_3$ are independently set to $0\pi$, $2\pi/3$, and $4\pi/3$, respectively. There are 9 markers in each plot showing the oscillation power detected through the directional coupler $P_{14}$ in dBm. The power of – 90 dBm corresponds to the noise level (no auto-oscillations). The On state corresponds to the power above -60 dBm. (D) All data combined.

**Figure 7:** Collection of experimental data obtained with a magnet placed on top of the ferrite film. (A) Schematics showing the position of the magnet on top of the ferrite film (B). Collection of experimental data showing the phase triplets that lead to the auto-oscillations where the power taken through the $P_{14}$ coupler is above -60 dBm. The red markers show the data obtained for the system without a magnet as in Fig.6(D). The blue markers show the data obtained with a magnet placed on top the ferrite film.

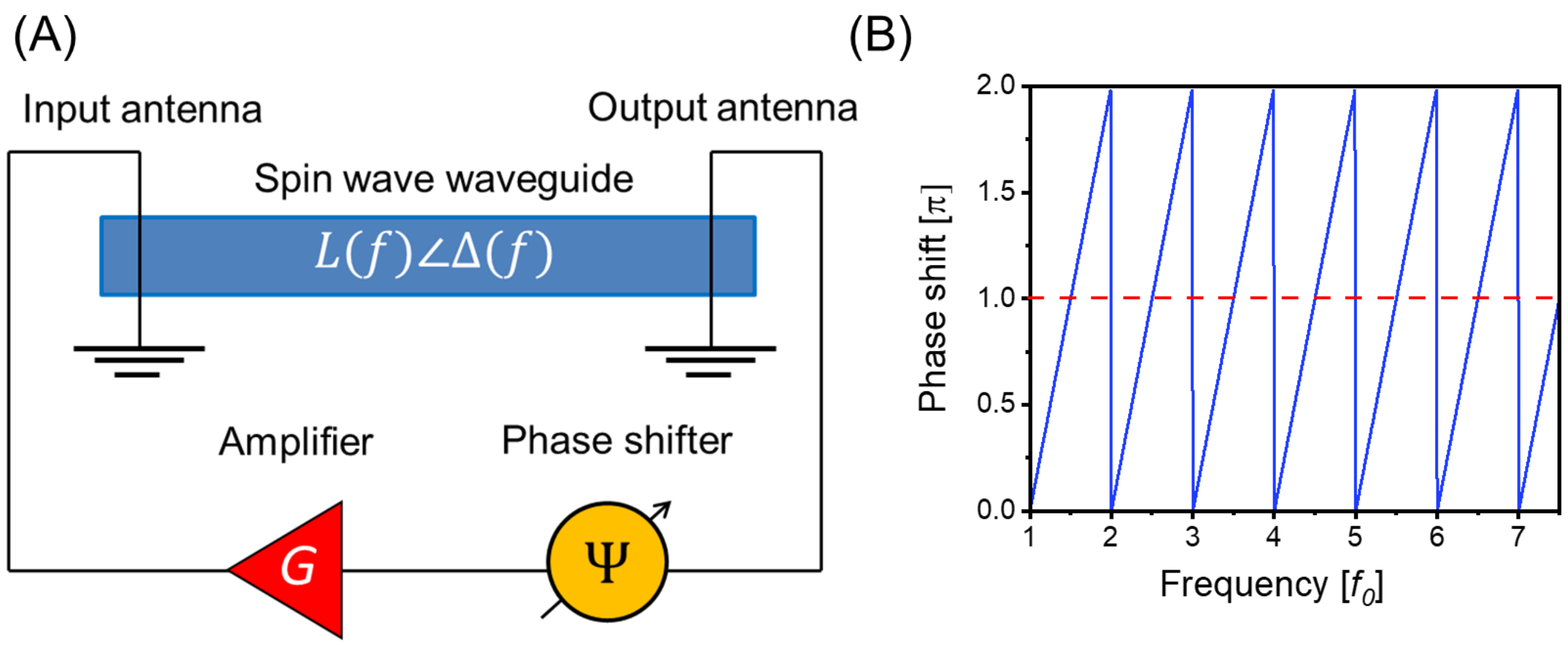
(A)
Input antenna
Output antenna
Spin wave waveguide
$L(f)\angle\Delta(f)$
Amplifier
Phase shifter
G
Ψ
(B)
Phase shift [π]
Frequency [$f_0$]
2.0
1.5
1.0
0.5
0.0
1
2
3
4
5
6
7

**Figure 1**

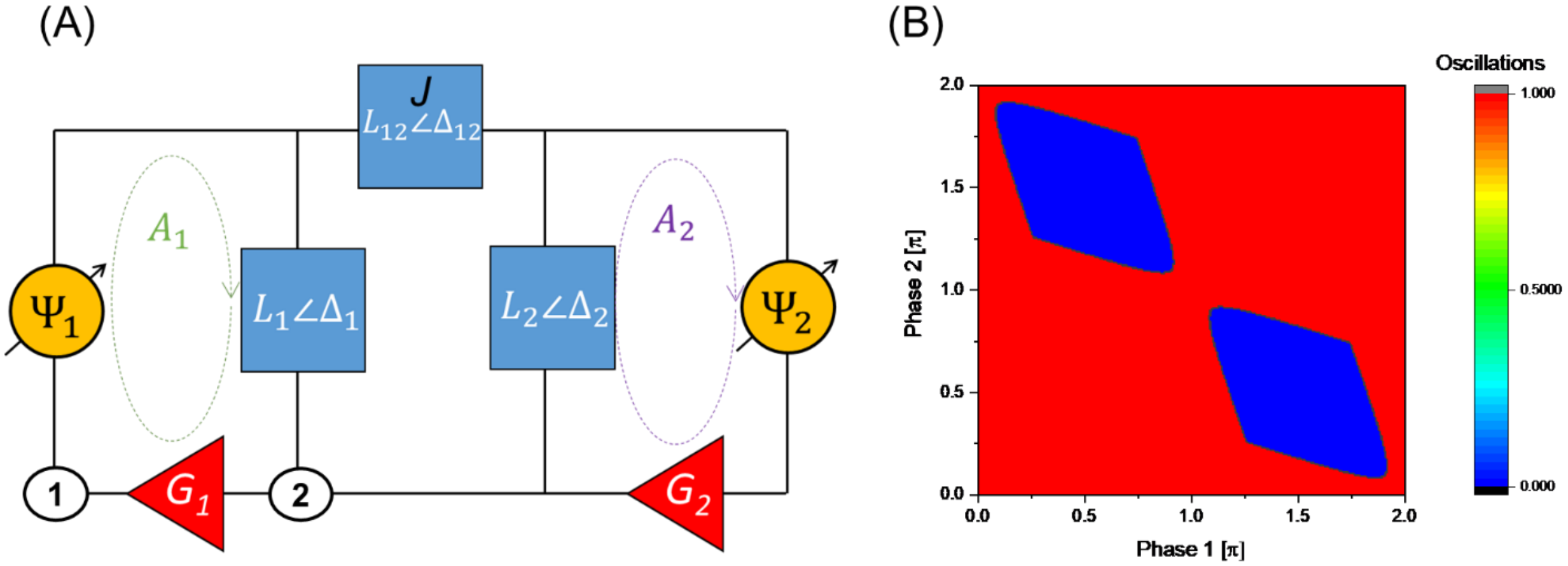
(A)
J
$L_{12}\angle\Delta_{12}$
$A_1$
$A_2$
$\Psi_1$
$\Psi_2$
$L_1\angle\Delta_1$
$L_2\angle\Delta_2$
1
2
$G_1$
$G_2$
(B)
Oscillations
1.000
0.5000
0.000
Phase 2 [π]
Phase 1 [π]
0.0
0.5
1.0
1.5
2.0

**Figure 2**

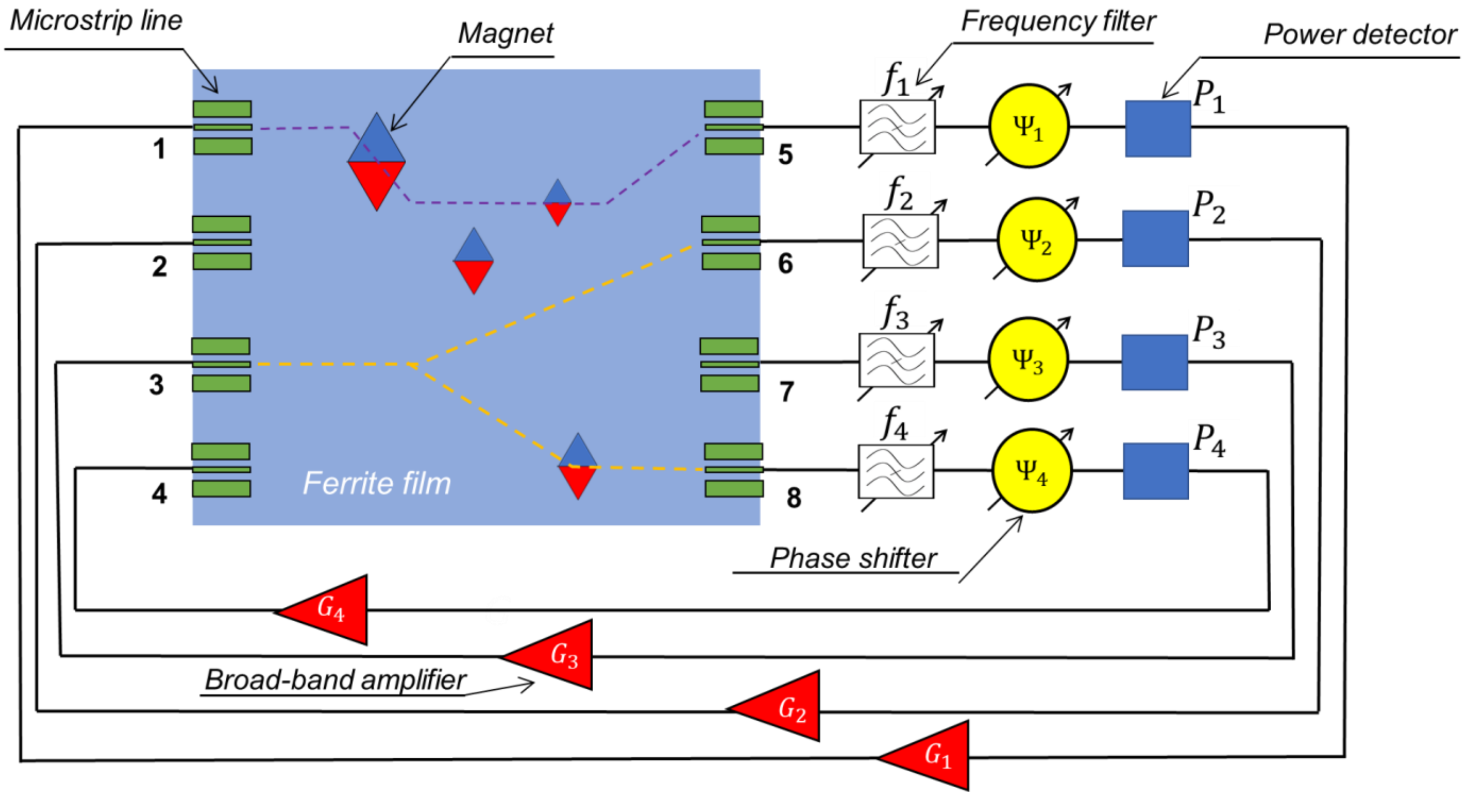
Microstrip line
Magnet
Frequency filter
Power detector
1
2
3
4
5
6
7
8
Ferrite film
$f_1$
$f_2$
$f_3$
$f_4$
$\Psi_1$
$\Psi_2$
$\Psi_3$
$\Psi_4$
$P_1$
$P_2$
$P_3$
$P_4$
Phase shifter
$G_4$
$G_3$
$G_2$
$G_1$
Broad-band amplifier

**Figure 3**

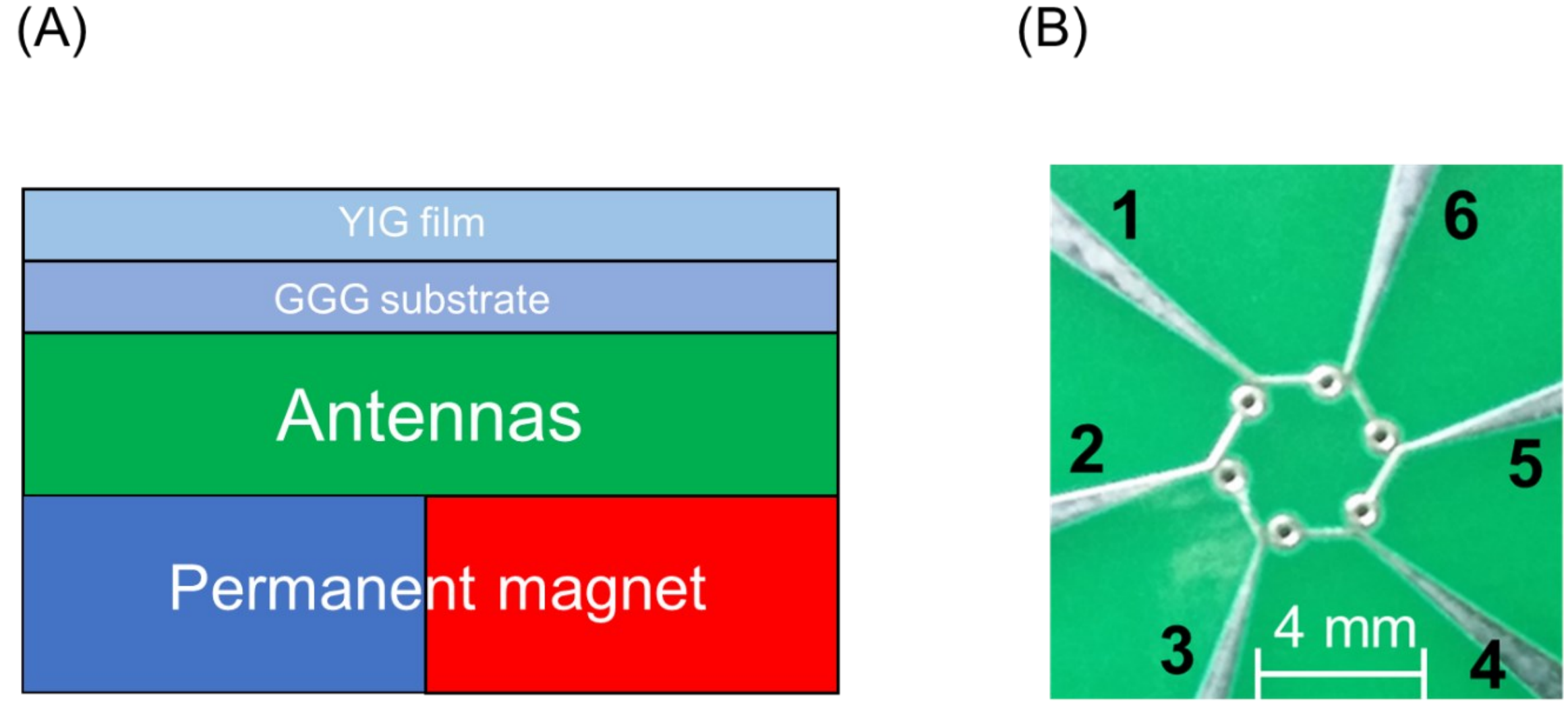
(A)
(B)
YIG film
GGG substrate
Antennas
Permanent magnet
1
6
2
5
3
4 mm
4

**Figure 4**

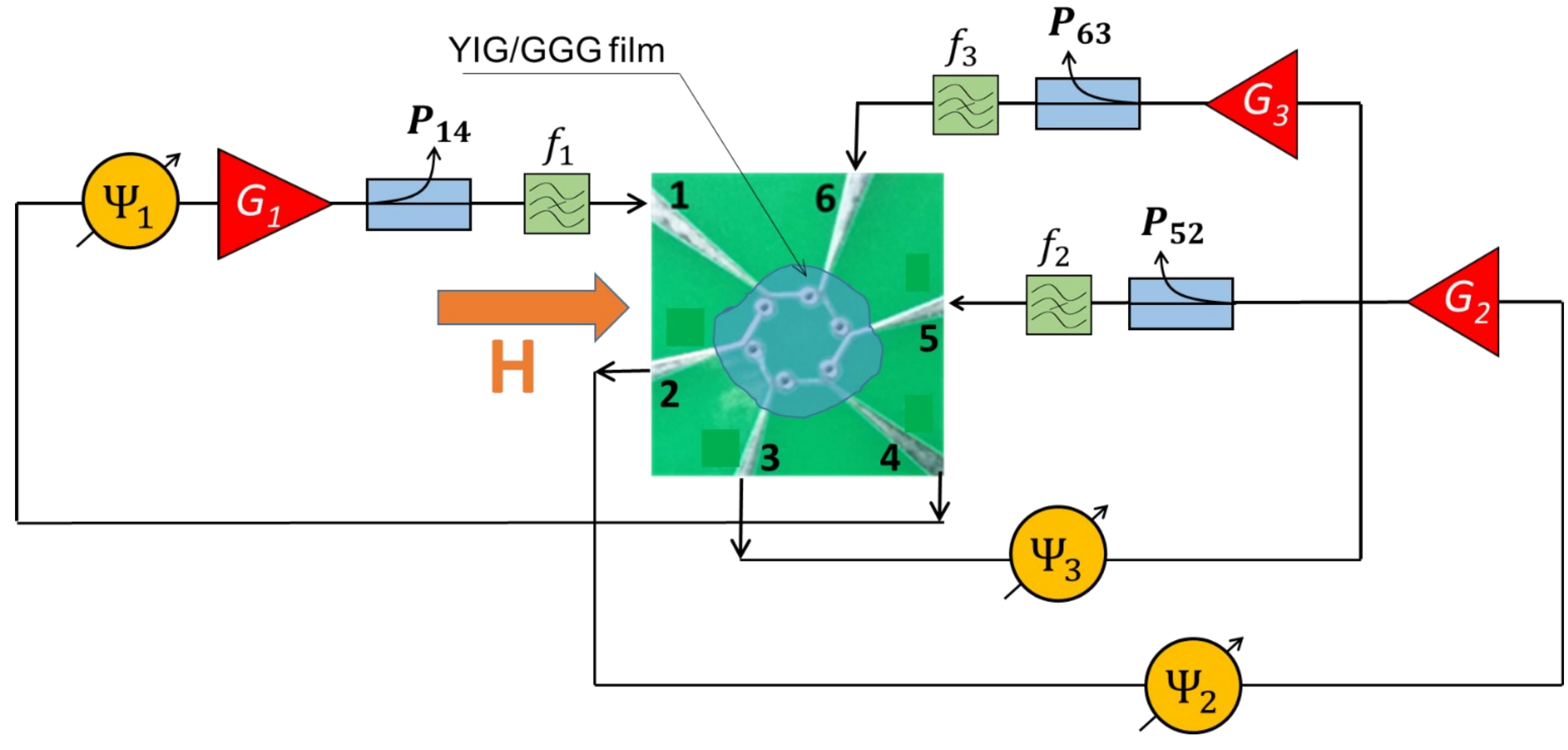
YIG/GGG film
$P_{63}$
$f_3$
$G_3$
$P_{14}$
$f_1$
$\Psi_1$
$G_1$
1
6
$f_2$
$P_{52}$
$G_2$
5
H
2
3
4
$\Psi_3$
$\Psi_2$

**Figure 5**

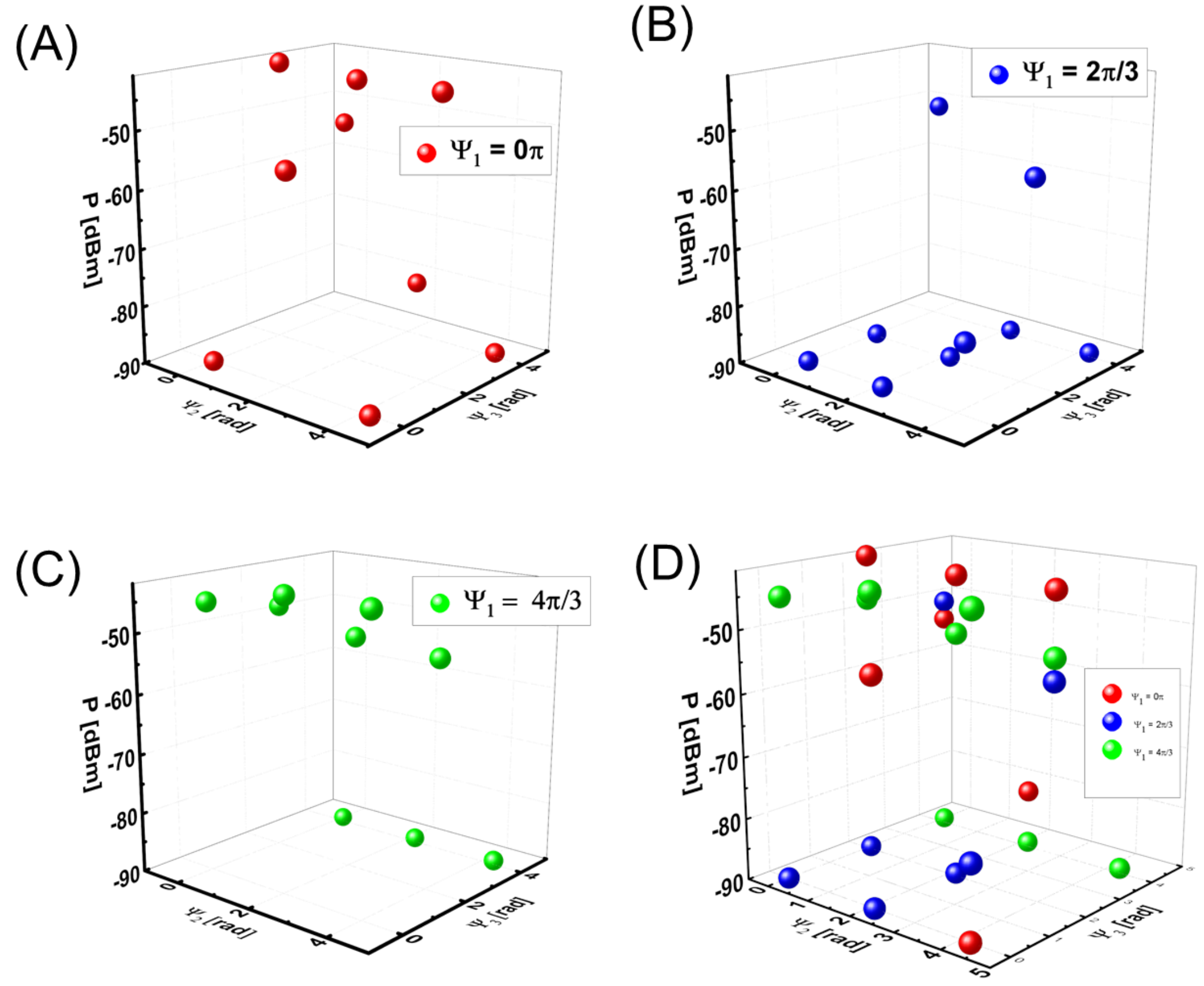

**Figure 6**

(A)

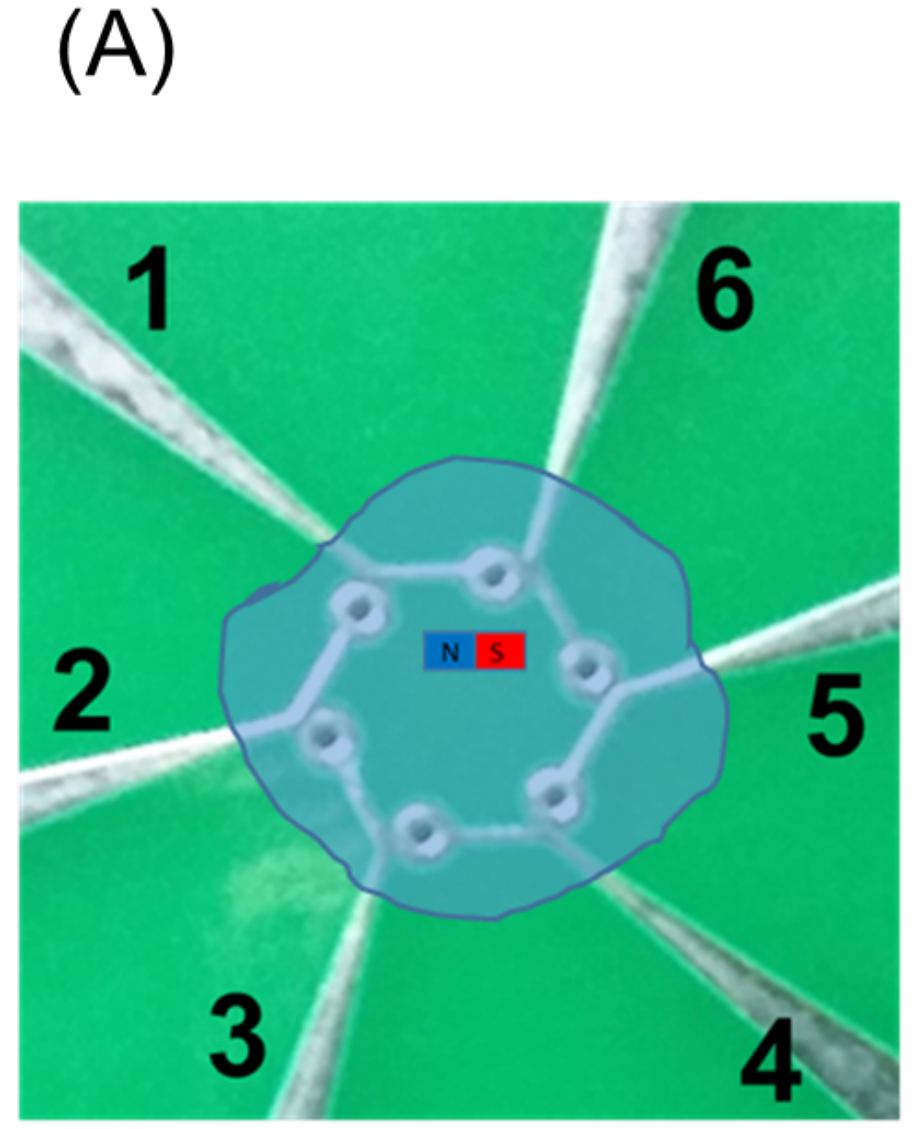
1
6
2
N
S
5
3
4

(B)

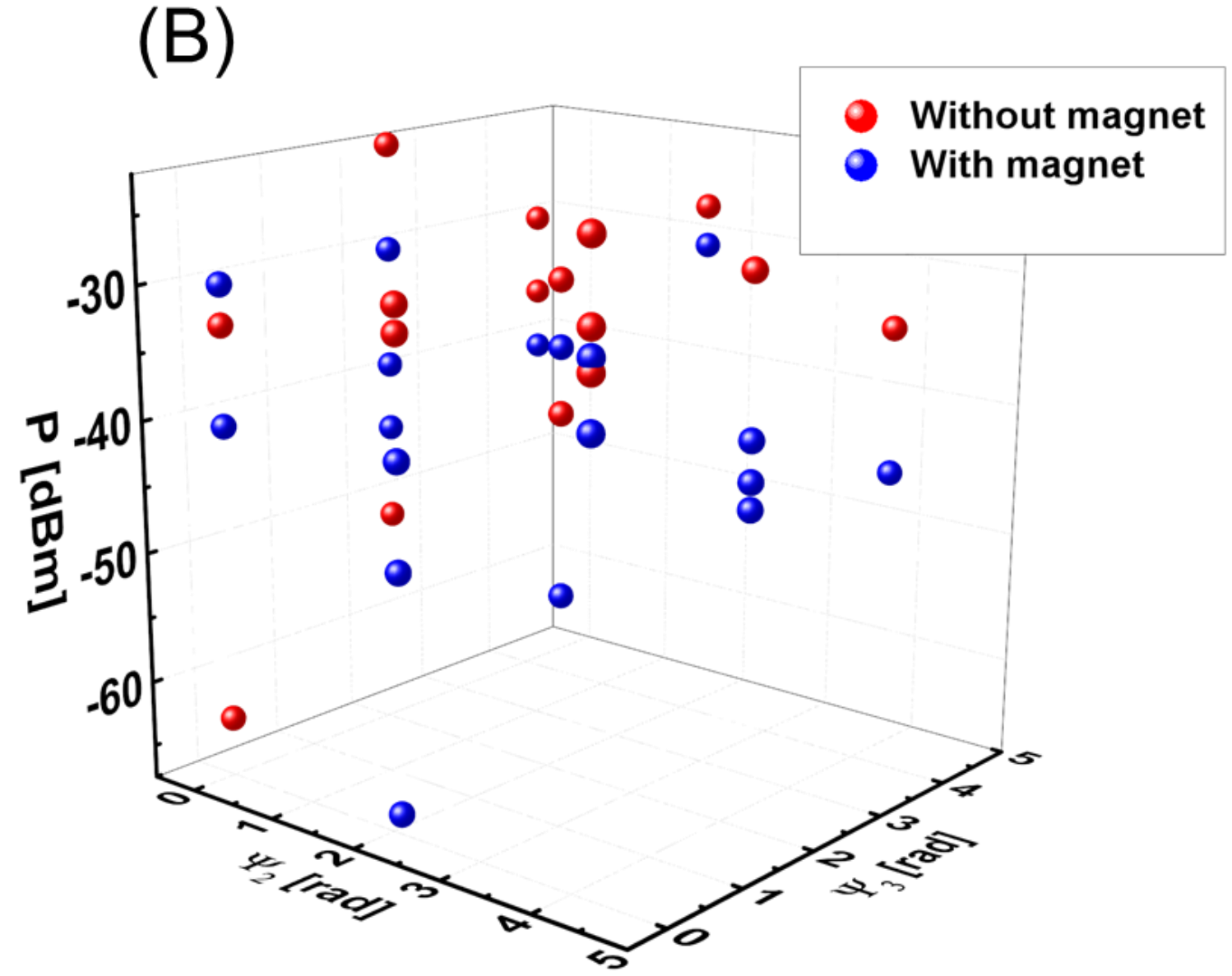
Without magnet
With magnet
P [dBm]
-30
-40
-50
-60
Ψ2 [rad]
Ψ3 [rad]
0
1
2
3
4
5

**Figure 7**